\documentclass[preprint,prd,showpacs,amsmath]{revtex4}

\usepackage{graphicx}
\usepackage[dvips]{epsfig}
\usepackage{dcolumn}
\usepackage{bm}

\usepackage{amsmath}
\usepackage{amssymb}

\begin{document}

\title{ No-thinning simulations of extensive air showers and small
  scale fluctuations at the ground level}

\author{
V.A.~Kuzmin$^1$ and G.I.~Rubtsov$^{1,2}$
}

\affiliation{$^1$Institute  for Nuclear Research of the Russian Academy of Sciences,
  Moscow 117312, Russia\\
$^2$Department of Physics, M.V.~Lomonosov Moscow State University, Moscow 119992, Russia}

\date{
February 16, 2007}

\begin{abstract}
The particle density in extensive air showers fluctuates at the ground
level. These fluctuations, at the scale of the scintillator detector
size (several meters), lead to the diversity of the individual
detector responses. Therefore, small scale fluctuations contribute to
the error in the estimation of the primary energy by a ground
array. This contribution is shown to be non-Gaussian. The impact on
the primary energy spectrum measured by a ground array is
estimated. Ir is argued that super-GZK events observed by AGASA
experiment do not result from the energy overestimation, due to small
scale fluctuations, of lower energy events.

\end{abstract}


\maketitle

\section{Introduction}

A significant part of conclusions on ultra-high energy cosmic
rays~(UHECRs) is made today on the basis of the quantities observed by
ground detector arrays. Typical ground detector registers only small
fraction $(<10^{-6})$ of the shower particles at the ground level. The
reading of the individual detector in the array is determined by the
local density of particles in the shower. The latter is affected by
the small scale fluctuations within the shower.

A typical EAS, induced by a vertical proton with a primary energy of
about $10^{20}$~eV contains about 20 billions of particles with
energies above $1$~MeV at the ground level.  The modelling of such a
huge number of particles is time and resource consuming
process. Monte-Carlo simulations of air showers induced by ultra-high
energy cosmic rays often involve procedures like
``thinning''~\cite{Hillas:1997tf}, aimed to reduce the effective
number of particles in a calculation.  These procedures make it
impossible to estimate small scale fluctuations in a reliable
way. Suggested ``unthinning'' procedure~\cite{unthinning} washes out
small scale fluctuations by its definition.

In the present work we use some artificial vertical proton-induced
air-showers with energies up to $10^{18}$~eV, simulated without
thinning, to estimate the impact of the small scale fluctuations on the
energy spectrum in the ultrahigh-energy region, observed by a ground
detector array with detectors simular to ones used in AGASA
experiment.

The fluctuations were earlier estimated experimantally by the Akeno
experiment~\cite{Teshima:1986rq}. The standard deviation value for the
fluctuations was obtaned but the fluctuation distribution has not been
discussed.

\section{Energy estimation procedure}

Ground array experiments traditionally use average density of
scintillator signal $S(r_{const})$, measured in units of vertical
equivalent muons~(VEMs), at a fixed constant core distance $r_{const}$
as the energy estimator. The $S(r_{const})$ is usually obtained by
fitting the detectors readings by the empirical lateral distribution
function~(LDF)~$S(r)$.

For example, AGASA experiment uses $r_{const}=600~m$ and the following
empirical LDF~\cite{AGASA:LDF},
\begin{equation}\label{agasaldf}
S(r)\propto\left( \frac{r}{R_{M}} \right)^{-1.2}
\left(1+\frac{r}{R_{M}} \right)^{-(\eta-1.2)}
\left(1+\left(\frac{r}{R_{1}}\right)^{2}
\right)^{-0.6},
\end{equation}
where
$$
\eta=3.97-1.79 (\sec\theta-1), ~~~
R_{M}=91.6~{\rm m}, ~~~
R_{1}=1000~{\rm m}.
$$ Only detectors with 300~m$\le r \le$ 1000~m are used for the
fit. Average scintillator detector responses for AGASA experiment are
published in~\cite{Sakaki}. To ensure fit quality, AGASA implements
the following procedure: the worst detector is excluded from the fit
in the case of bad $\chi^2$ ($\chi^2/N > 1.5$)~\cite{TakedaThesis} and
the procedure is repeated until $\chi^2/N \le 1.5$. Detectors
experiencing largest local fluctuations are expected to be excluded by
this procedure.

\section{Simulations}

We have generated several showers without thinning with primary
energies ranging from $10^{17}$ to $10^{18}$~eV. The simulations were
preformed with CORSIKA~\cite{CORSIKA}.  QGSJECT~01c~\cite{QGSJET-1},
QGSJET~II~\cite{QGSJET}, GHEISHA~\cite{GHEISHA} and
EGS4~\cite{Nelson:1985ec} models were used in simulations. The
simulations without thinning are time consuming and CPU-time and
storage required grows nearly linearly with primary energy. This is
why we limit our simulations up to energies of~$10^{18}$~eV.

All datafiles are made publicly available within the public library of
artificial air showers called ``Livni'', so all the results of this
and the following works may be confirmed using the same dataset.  The
library may be used for any other studies of the structure of air
showers. The detailed information on the library content and access
rules are available at the website~\cite{livni}. Hereafter we use
references to the library showers in the form of livni:codename,
e.g. reference to the shower, named in the library as ``18-3'' will be
livni:18-3.

\section{Results}

In the present work we intend to study fluctuations in air showers
induced by the highest energy primary particles. Unfortunately,
simulations of artificial air showers of such a high energy is
impractical. In order to make statements on $10^{20}$~eV showers we
consider scintillator detectors with an area 100 times larger than
normal detector area ($2.2~m^2$ in AGASA) and utilize simulated
showers with a primary energy of $10^{18}$~eV. This of course make our
analysis approximate. The procedure is justified by the fact that the
lateral distribution functions of scintilation signal density for
energies $10^{18}$~eV and $10^{20}$ have a similar
shape~\cite{AGASA:LDF}. Furthermore, we have calculated the
cross-correlation function, $C_S(\vec d) = \frac{\int d\vec r S(\vec
r)~S(\vec r+\vec d)}{\int d\vec r S^2(\vec r)}$, of scintillation
density at the ground level. We found that cross-correlation function
is close to zero on the scale of the detector size, $|\vec
d|\gtrsim0.4~m$.

\begin{figure}[htb]
\includegraphics[angle=270,width=0.9\columnwidth]{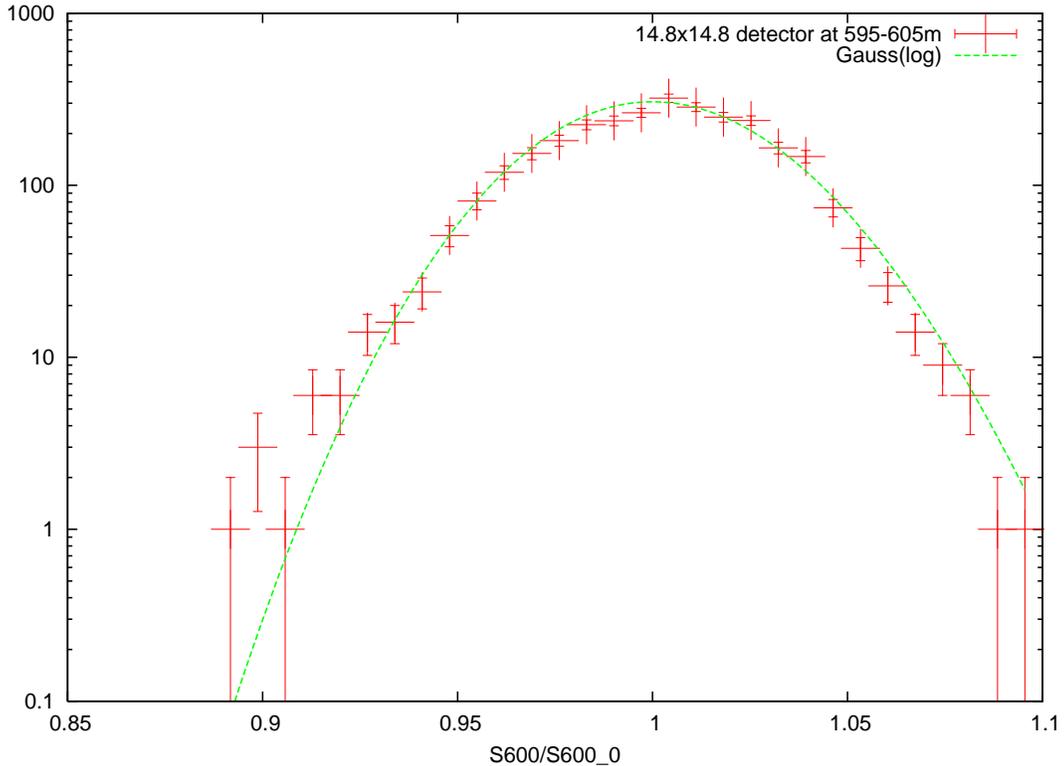}
\caption{\label{single}Scintillation density distribution over
  detectors (size of $14.8~m\times14.8~m$) centered at [595~m;605~m]
  core distance. Horizontal axis: scintillator density, normalized to
  average, Vertical axis: number of detectors with a signal in a bin
  centered in $S(600)$ (livni:18-3 shower).}
\end{figure}

The scintillation density distribution over detectors centered at the
core distance between 595 and 605 meters is shown in
Fig.~\ref{single}. To produce this plot we assume the ground to be
fully covered by the detectors.  It can be seen that an individual
detector may be exposed to a larger or smaller density than an average
one and the central part of the distribution obeys the Gaussian law in
$\log(S)$ scale. Let us note that the plot refers to 600~m core
distance where the fluctuations are small. At larger distances an
individual detector may be exposed to up to 100 times larger particle
density than an average, though the probability is small.

In order to reconstruct the observables we assume the ground array
detector to consist of 100 plastic 5 cm thick scintillators ($14.8~m$
$\times$ $14.8~m$) forming square lattice covering the area of
$100~km^2$. Our intent is to make our detection procedure close to one
used for the analysis of AGASA experimental
data~\cite{agasares,AGASA_Eest}. The detector responses are estimated
using AGASA average detector response functions~\cite{Sakaki}.  Each
simulated shower has been detected 30000 times with different core
positions within the ground array and azimuthal angles with respect to
the array.

Fitting responses of the detectors at core distances from 300 to 1000
meters with the AGASA experimental LDF~Eq.~(\ref{agasaldf}), we obtain
$S(600)$. Following the AGASA procedure to ensure fit quality we
exclude the worst detector from the fit in the case of bad $\chi^2$
($\chi^2/N > 1.5$) \cite{TakedaThesis}.  The distribution of the
number of excluded detectors is presented in Table~1~\footnote{It
should be noted that the distribution presented here is not
neccesarily the same as in the analysis of the original AGASA data, as
our definition of $\chi^2$ does not include detector fluctuations and
therefore is different from the experimental one.}.
\begin{table}[htb]
\begin{center}
\begin{tabular}{|c|c|}
\hline
Number of excluded & $\%$ of cases \\
detectors & ~\\
\hline
1            & $23\%$ \\
2 & $15\%$\\
3 & $9\%$\\
4 & $7\%$\\
$\ge 5$ & $7\%$ \\ 
\hline
\end{tabular}
\end{center}
\caption{ The distribution of the number of excluded detectors in the
fit quality assurance procedure~(livni:18-3 shower)
}
\end{table}

\begin{figure}[htb] 
\includegraphics[angle=270,width=0.9\columnwidth]{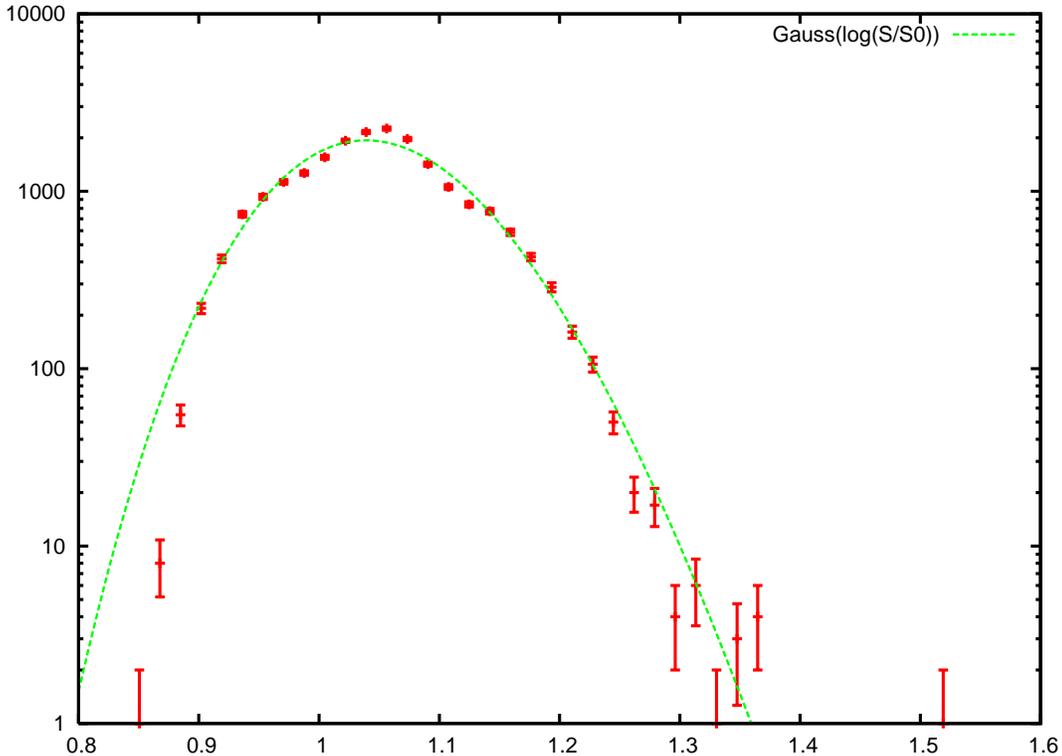} 
\caption{\label{sdistr}Distribution of reconstructed S(600) normalized
  to the average value (livni:18-3 shower). Estimated energy
  distribution has the same form as an estimated energy is nearly
  proportional to $S(600)$.}
\end{figure}
The resulting distribution of reconstructed $S(600)$ calculcated for
one artificial air shower is shown on Fig.~\ref{sdistr}. The
reconstruction error may depend on the first interaction producing a
shower, as we discuss later in Sec.~\ref{sec:lim}. The main part of
the distribution may be fit with the Gaussian in $\log(S)$ scale. The
same type of profile is suggested by AGASA for $S(600)$ experimental
error distribution~\cite{AGASA_Eest}. Finally, we estimate one-sigma
error for $S(600)$ reconstruction for $10^{20}$~eV air showers due to
small scale fluctuations as $7\%$. In rare cases $S(600)$ may be
overesimated by factor of $1.5$, hovewer, the probabilty of this is
less than $10^{-4}$. We may also see that the part of the distribution
which corresponds to energy overestimation is broader than the
underestimation part. The estimate does not include the fluctuations
of the detector response, which are present in the experiment and are
of the same order of magnitude~\cite{Teshima:1986rq}.

\begin{figure}[htb] 
\includegraphics[angle=270,width=0.9\columnwidth]{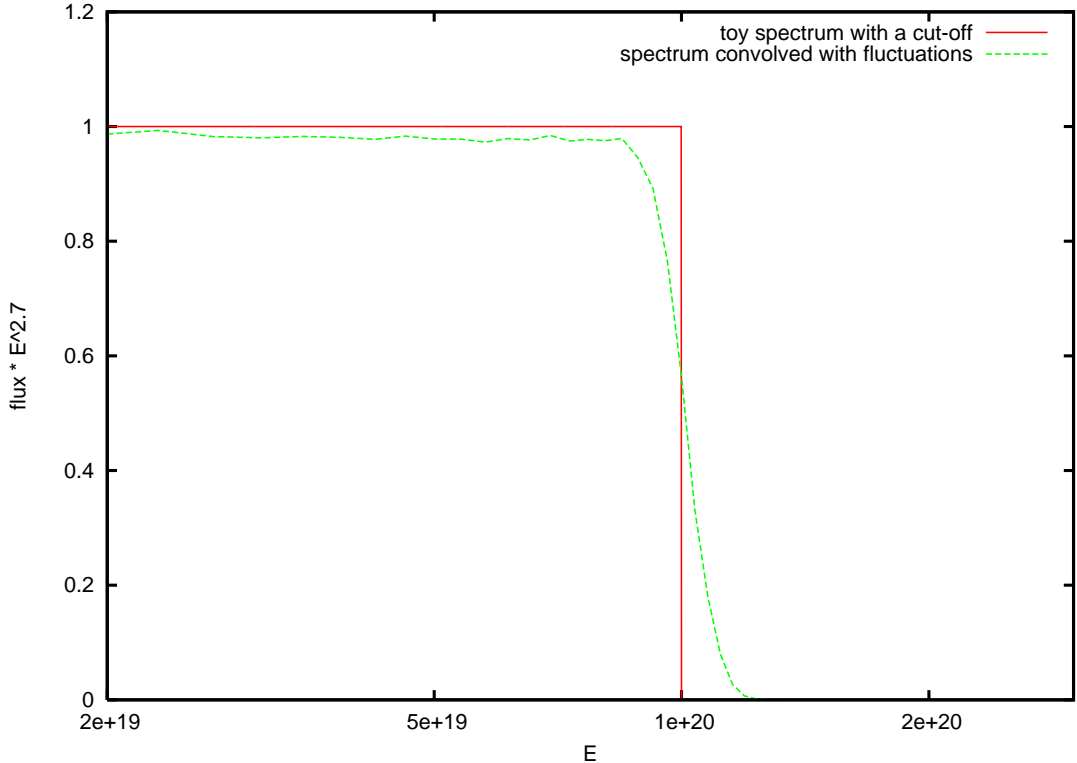} 
\caption{\label{spec}The impact of the small scale fluctuations on the
  toy primary spectrum with a cut-off. Dashed line shows the
  convolution of a toy spectrum and an energy estimation error
  presented in Fig.~\ref{sdistr}}
\end{figure}

The fluctuations discussed above may affect the primary spectrum
observed by a ground array. Significant energy overestimation even in
a relatively small number of cases may influence the experimental
conclusion on the presence of the GZK cut-off~\cite{G,ZK}. Let us
assume a toy primary spectrum with a spectral density proportional to
$E^{-\alpha}$, $\alpha = 2.7$ up to the energy of $10^{20}$~eV and
equal to zero for higher energies. We have calculated the convolution
of our toy spectrum and energy estimation fluctuation distribution
(which is the same as $S(600)$ fluctuation distribution, presented in
Fig.~\ref{sdistr}). The resulting spectrum is shown in
Fig.~\ref{spec}. We see that the fluctuation contribution to the
spectrum may be considered minor for the GZK-predictions: the
probability of energy overestimation by a factor of $1.5$ is less
than~$10^{-4}$. We conclude that super-GZK events observed by AGASA
experiment do not result from energy overestimation, due to small
scale fluctuations, of lower energy events.

\section{Limitations}
\label{sec:lim}
One limitation of the above analysis is that is does not include the
fluctuations of the detector response which are of the same order of
magnitude~\cite{Teshima:1986rq}. The fluctuations within the detector
may be accounted for by combining the library showers with relevant
Monte-Carlo simulations of the detector.

The second problem one should care about is that the magnitude of
small-scale fluctuations may be different for showers having different
first interactions. To estimate this effect we analysed 20 showers
with primary energy of $10^{17}$~eV and 3 showers with primary energy
of $10^{18}$~eV. For each shower we have estimated $\alpha(r) =
\sigma^2(r)/\bar S(r)$, where $\sigma$ is a standard deviation of the
detector response, measured in VEMs, at distance $r$~(calculated for
the ensemble of detectors centered at core distances close to $r$) and
$\bar S(r)$ is the average detector response. As long as the
correlation function $C_S$ is zero $\alpha(r)$ doesn't depend on the
detector size. This motivates the choice of $\alpha(r)$ as an
estimator for the fluctuation magnitude in a particular artificial
shower. For pure Poisson distribution (which would be the case for
equivalent independent particles) $\alpha \equiv 1$. Average and
maximum values of $\alpha(600)$ for $10^{17}$~eV showers are 0.47 and
0.51, respectively, with a standard deviation of $7\%$.  Three studied
$10^{18}$ showers have $\alpha(600)$ equal to 0.48, 0.57 and 0.69.
These numbers imply that the difference between statistical properties
of different showers is not very substantial. To interpret these data
further, we assume that the character of the fluctuations does not
change dramatically when the primary energy changes from $10^{18}$ to
$10^{20}$~eV. As the study is based on a small number of artificial
showers, we also have to assume practical inexistence of air showers
with extremely large fluctuations. The first assumption may be checked
by a simulation of $10^{20}$~eV artificial air shower without thinning
and the second by simulating hundreds of showers without
thinning. Both simulations are extremely resource consuming and yet
are expected to be possible in the nearest future.

Thus, the difference in the fluctuation magnitude in different studied artificial
showers is not substancial for the conclusions of the work, although
our study is limited by the following:
\begin{itemize}
\item The primary energy of artificial showers does not exceed
$10^{18}$~eV.
\item The number of artificial events studied is relatively small and
  does not exclude the existence of paricular air showers with
  extremely large fluctuations in them.
\end{itemize}
The above limitations may be checked in the future.

\section{Conclusions}

We calculated the cross-correlation function of the scintillation
signal density and found it to be close to zero at the detector
scale~$\gtrsim~0.4m$. We estimated a contibution of small scale
fluctuations on the detector scale to the energy reconstruction error
by a ground array at the level of about $7\%$ for primary energy of
about $10^{20}$~eV. The contribution, although found to be
non-Gaussian, is minor for GZK predictions. The study, however, has
certain limitations; we discussed ways to get rid of them.

{\bf Acknowledgements.}

We are indebted to L.~G.~Dedenko, D.~S.~Gorbunov, M.~Kachelrie\ss,
A.~Kusenko, R.~Peccei, C.~Rebbi, V.~A.~Rubakov, S.~Shinozaki,
M.~Teshima, G.~Thomson, I.~I.~Tkachev, and S.~.V~Troitsky for useful
discussions and criticism. G.R. is grateful to the Boston University,
where the work has been started.

This work was supported in part by the INTAS grant 03-51-5112, by the
Russian Foundation of Basic Research grant 05-02-17363 (GR), by the
grants of the President of the Russian Federation NS-7293.2006.2
(government contract 02.445.11.7370), by the fellowships of the
"Dynasty" foundation (awarded by the Scientific Council of ICFPM,
GR). Numerical part of the work was performed at the computer cluster
of the Theoretical Division of INR RAS.


\end{document}